# Ferroelectricity and Antiferroelectricity in Elemental Group-V Monolayer Materials

Chengcheng Xiao[1], Fang Wang[1], Shengyuan A. Yang[2], Yunhao Lu[1]*

[1] State Key Laboratory of Silicon Materials, School of Materials Science and Engineering, Zhejiang University, Hangzhou, 310027, China

[2] Research Laboratory for Quantum Materials, Singapore University of Technology and Design, Singapore 487372, Singapore

Abstract

Ferroelectricity is usually found in compound materials composed by different elements. Here, based on first-principles calculations, we reveal the first example of spontaneous electrical polarization and ferroelectricity in stable two-dimensional elemental materials: elemental Group-V (As, Sb, and Bi) monolayers. The polarization is due to the spontaneous lattice distortion with atomic layer buckling. Interestingly, for Bi monolayer, apart from the ferroelectric phase, we find that it can also host an antiferroelectric phase. The Curie temperatures of these elemental materials can be higher than room temperature, making them promising for realizing ultrathin ferroelectric devices of broad interest.

Ferroelectric materials, especially with reduced dimensions, have attracted significant interest in recent years due to their potential applications for high-density nonvolatile memories [1], sensors, or nanoscale electronic devices [2]. For the perovskite ferroelectrics (e.g., $PbTiO_3$ and $BaTiO_3$), the transition temperature ($T_C$) usually decreases as the material thickness is reduced, and a minimum thickness of a few unit cells is required to maintain the spontaneous polarization at room temperature [3-5]. Very recently, two-dimensional (2D) ferroelectricity (FE) above room temperature has been predicted in Group-IV monochalcogenides atomic monolayers, and has been experimentally demonstrated in SnTe monolayer [6]. This FE behavior originates from the in-plane lattice distortion related to the less-ionic or resonance bonding between group-IV and group-VI atoms [7, 8].

All reported ferroelectrics, to our knowledge, are compounds composed by different elements; whereas no intrinsic ferroelectricity is ever reported for elemental materials. Is it possible to have elemental ferroelectrics? This fundamental question has not been addressed yet. Although quite counterintuitive, we note that the appearance of FE in elemental materials does not contradict the fundamental requirement of broken centrosymmetry. The major challenge lies in that most elemental materials are stabilized in non-polar structures with high symmetry, forbidding an electric polarization. One needs a strong instability to destroy the centrosymmetry and drive the system towards a polarized state.

In this work, we answer the above question by predicting robust FE in a family of stable 2D elemental materials — the monolayers of group-V elements As, Sb, and Bi. We show that these materials are stabilized in one of the two degenerate structures, which may be viewed as distorted phosphorene lattice structures. The distortion breaks centrosymmetry and generates a spontaneous electric polarization. The FE phase is robust, with polarization values on the order of $10^{-10}$ C/m, and the Curie temperatures estimated from both Monte Carlo simulations and *ab-initio* molecular dynamics simulations are above room temperature. Interestingly, for Bi monolayer, we also find a metastable antiferroelectric (AFE) phase. The displacive instabilities towards FE phase (and also AFE phase for Bi) can be related to the soft optical modes in the phonon spectrum of the undistorted structure. Our result reveals a fundamentally new phenomenon and offers a realistic platform to explore the intriguing physics of 2D FE and AFE phases as well as promising device applications.

The group-V elements can crystalize in a variety of structures. In 2D form, the most well-known structure of P, the lightest element of the group occurring in solid form, is the phosphorene structure (space group *Pmna*). As illustrated in Fig.1(a) and Fig.1(b) (of Phase B), the monolayer consists of two flat atomic layers, with four atoms in a primitive unit cell. Importantly, the structure preserves an inversion center, hence forbidding FE. FE may be induced by extrinsic means, e.g., via an external electric field[9], or by substituting different elements for P on the two sublattices of phosphorene to break centrosymmetry, as is the case in previous reports on 2D FE in SnTe-family materials[10, 11]. Distinct from these mechanism, the As, Sb, and Bi monolayers break centrosymmetry in a different way. In the 2D limit, these latter group-V elements can all be stabilized in a structure similar to phosphorene, which has been experimentally demonstrated for Sb[12] and Bi[13, 14]. A crucial difference is that now each atomic layer becomes buckled due to the reduced $sp^3$ hybridization as compared with P. This buckling has been confirmed by experiment on Bi monolayer[13], and was shown to be the key factor in determining its topological property[14]. Here, we note that the buckling also breaks the centrosymmetry, which is the necessary ingredient for a nonzero polarization.

To study the structural and electronic properties, we performed first-principles calculations based on the density functional theory (DFT). The calculation details are presented in the Supplemental Material. The calculation confirms the distorted phosphorene lattice structure (space group *Pmn2₁*) for monolayer As, Sb, and Bi. The detailed structural parameters are shown in Table S2. Of most importance is the finite buckling in the obtained structure. In Fig.1(a) and 1(b), two of the four basis atoms (blue-colored) are considered as fixed, and the buckling can be viewed as resulting from the vertical shift of the other two (red-colored) atoms. The buckling may be quantified by the two heights $h_i = z_R^i - z_B^i$ (see Fig.1(b)), where $i(=U,L)$ denotes the two atomic layers (upper or lower), and the subscripts $R$ and $B$ denote the red and blue colored atomic sites, respectively. One notes that when $h_U = h_L = 0$ (denoted as phase B), the structure converts back to the phosphorene structure (*Pmna*) with preserved centrosymmetry. For As, Sb, and Bi monolayers, $h_i$'s are nonzero, and there are two stable degenerate structures with $h_U = h_L > 0$ (denoted as phase A) or $h_U = h_L < 0$ (denoted as phase A'), as illustrated in Fig.1(b). Both structures break centrosymmetry, and they are related to each other via an inversion operation. Hence, if A phase has a finite polarization *P*, A' phase must have the opposite polarization of −*P*. Taking As monolayer as an example, the free-energy contour obtained by first-principle calculations is plotted in Fig.1(c), which shows that the two stable structures A and A' are connected through a saddle point corresponding to B (with an energy difference of 1.46 meV/atom), and feature a characteristic anharmonic double-well potential (see Fig.S3(c)). This observation strongly hints the existence of FE.

The finite FE is confirmed by our DFT calculations. Using the modern Berry phase method, we find that A and A' phases have significant and reversed spontaneous polarizations along the in-plane *x*-direction. Note that the polarization is constrained to be along the *x*-direction due to the presence of mirror or glide mirror symmetries in the other two directions. The values of the spontaneous polarization at zero temperature ($P_S$) are listed in Table I. The value increases with the atomic number. $P_S$ of Bi monolayer is even larger than $1.5 \times 10^{-10}$ C/m, which is comparable to those in group-IV monochalcogenides.

In bulk ferroelectrics, the displacive instability that drives the FE transition is related to the soft optical modes in the phonon spectra of undistorted structures. As the temperature drops below the phase transition temperature, the soft optical modes become "frozen" as their frequencies become imaginary, eventually causing the structural distortion. In Fig.2, we plot the phonon spectra for the monolayers of group-V elements with the *undistorted* phosphorene structure (i.e., with $h_U = h_L = 0$). One observes that only the spectrum of P is free of any soft phonon modes throughout the Brillouin zone. For all the other elements As, Sb, and Bi, there appear pronounced soft optical modes at the Brillouin zone center Γ (denoted as $\lambda_1$). From analysis of the phonon eigenvectors, this imaginary mode at $\lambda_1$ corresponds to the motion of neighboring atoms of the same atomic plane in opposite directions along *z* with $h_U = h_L$, which results in the buckled structure. Indeed, with the distortion of buckling, the As, Sb, and Bi monolayers become dynamically stable, without any soft mode across the Brillouin zone in their phonon spectra (see Fig.S1).

Interestingly, in Fig.2(d) for Bi, besides the soft mode at Γ point, there are two additional soft optical modes: $\lambda_2$ between Γ and M, and $\lambda_3$ with the primary soft mode at the M point (0.5, 0.5, 0). The $\lambda_3$ mode corresponds to the rotation of tetrahedra (see Fig.S7(a)), similar to the rotation of octahedra in perovskite ferroelectrics. The $\lambda_2$ mode is more interesting. Analysis shows that it corresponds to a motion that the atoms in neighboring unit cells buckle in opposite directions (see Fig.S7). This will drive the system towards the configuration shown in Fig.3(a) and 3(b), which is a structure with preserved centrosymmetry. Here the phase is constructed using a 2×2 supercell. In Fig. 3(a) and 3(b), the red (green) colored atomic sites are displaced in +z (-z) direction relative to the blue colored sites, with $h_U^{(R)} = h_L^{(R)} = -h_U^{(G)} = -h_L^{(G)} = 0.67$ Å, where the additional superscripts *R* and *G* denote the red and green colored sites in the supercell. Thus, this kind of distortion produces a checkerboard pattern: the buckling displacements in neighboring unit cells are opposite. By our previous analysis, this will generate opposite polarization between neighboring cells, leading to an AFE phase. The calculated local polarizations confirm this picture, and the magnitude of polarization in each unit cell is about 1.5×10$^{-10}$ C/m. In Fig.3(d), we plot the DFT calculated free-energy contours for the Bi monolayer with two parameters $h^{(R)}$ and $h^{(G)}$ (the displacements of upper and lower layers are taken to be the same, i.e., $h_U^{(R/G)} = h_L^{(R/G)}$, hence the subscripts are dropped). One indeed observes that besides the two FE minima of A and A' phase, there are two additional local minima C and C' corresponding to the AFE phase. The phonon calculation also confirms that this AFE structure is dynamically stable without any soft mode across the Brillouin zone (see Fig.S1). The AFE phase is higher in energy than the FE phase by 17 meV/atom. Note that the $\lambda_2$ and $\lambda_3$ soft modes do not appear for As and Sb. Our DFT calculation also verifies that they are not stabilized in the AFE structure like Bi. For example, Fig.3(c) shows the result for As monolayer, in which one only observes the two FE phases whereas the AFE phase does not appear.

The aforementioned soft optical modes capture the displacive instabilities that drive the system towards the FE phase at zero temperature. Meanwhile, it is also important to understand how the polarization behaves at elevated temperatures. Particularly, the Curie temperature $T_C$, the temperature above which the polarized is suppressed, is a crucial factor for practical applications of ferroelectrics. Here we adopt two approaches to estimate the Curie temperature: the Monte Carlo method and the *ab-initio* molecular dynamics (MD) simulation.

In the Monte Carlo approach, we used a Landau theory with the polarization $P$ as the order parameter. We first need to map the $(h_U, h_L)$ free-energy surface (in Fig.1(c)) as a function of $P$. Direct mapping is a computationally formidable task. Fortunately, one observes steep gradients traverse to the diagonal line with $h_U = h_L = h$ in Fig.1(c). Physically, this means the system prefers the "$h$-covariant" states which keep the bond length between the two red-colored atoms more or less unchanged. Hence, we may consider the mapping only for the 1D subset of configurations for these $h$-covariant states. From DFT calculations, we can then connect the free energy, the height $h$, and the order parameter $P$. Next, we construct a Landau-Ginzburg-type expansion of the energy in terms of the polarization $P_i$ of each unit cell:

$$E = \sum_i [\frac{A}{2}(P_i^2) + \frac{B}{4}(P_i^4) + \frac{C}{6}(P_i^6)] + \frac{D}{2}\sum_{\langle i,j \rangle}(P_i - P_j)^2 + \frac{E}{2}\sum_{\langle\langle i,j \rangle\rangle}(P_i - P_j)^2, \quad (\text{Eq.1})$$

where $i$ and $j$ label the unit cells, $\langle \ \rangle$ denotes the nearest-neighbor, $\langle\langle \ \rangle\rangle$ denotes the second-neighbor, coefficients *A, B, C, D*, and *E* are expansion parameters to be fitted by DFT calculations. The first three terms are associated with the energy contributed by each unit cell and they well describe the anharmonic double-well potential of our system (see Fig.S4). The last two terms are of the interaction energy between neighboring cells, which include the 2D geometrical character and are crucial for the ordering and the phase transition. We use parabolic function to describe this interaction, because our calculation indicates a harmonic relationship (see Fig.S8(c)). For simplicity, the polarization $P_i$ is constrained to be along *x*-direction, like in the previous treatment[11]. The model parameters obtained from fitting are shown in Table I. One observes that for these materials, the second neighbor coefficients are also sizable and cannot be neglected. We also checked the third and the fourth neighbor couplings, and found that they are orders of magnitude smaller hence can be discarded. Compared with As and Sb, Bi monolayer has a smaller first neighbor coefficient, which is connected to the appearance of AFE in Bi.

With this model, we can then proceed to investigate the temperature effects and the phase transition using Monte Carlo simulations. As shown in Fig.4, the estimated Curie temperatures for all three materials (Sb result shown in Fig. S8(a)) are above the room temperature (>300 K), showing that their FE orderings are robust and should be readily detectable in experiment. We adopted the heuristic form with $P(T) = 0$ for $T > T_C$; and $P(T) = \mu(T_C - T)^\delta$ (Eq.2) for $T < T_C$ to fit the result from Monte Carlo simulations, where $\mu$ is a constant and $\delta$ is the critical exponent. The fitted Curie temperatures and the model parameters are listed in Table II. The estimated Curie temperatures can be up to 680 K for Sb monolayers (Fig.S8(a)). Comparing the results in Fig.4(a) and 4(b), one also observes that the decrease of polarization with temperature for Bi is smoother than that of As, which may be attributed to the metastable AFE phases existing for Bi. At elevated temperatures, the system could first develop AFE domains that coexist with FE domains before all correlations are destroyed at higher temperatures.

These results are verified by the *ab-initio* MD simulations. In the MD approach, the temperature, the total energy, and the average polarization can be tracked in the simulation, without any fitting parameters. The MD results are also plotted in Fig.4. One observes a fairly good agreement between the results from two approaches. This confirms that the FE in these materials is robust and can persist above room temperature. It also suggests that the simple effective model (1) captures the essential physics.

Ferroelectric materials must also exhibit pyroelectric response, which is given by the derivative $dP/dT$. In the lower panels of Fig.4(a) and (b), we plot the average pyroelectric response versus temperature using the Monte Carlo results. The dip is shallow for Bi and deep for As (and Sb). The sharp dips for As and Sb should be detectable in experiment, when the temperature is swept through the Curie temperature.

A few remarks are in order before closing. First of all, this is the first time that robust FE is revealed in stable 2D elemental materials. Given that Sb and Bi monolayers with the studied structure have already been realized in experiment[14, 15], we expect that their FE properties should be readily verified in the near future. The metastable AFE phase for Bi is also reported here for the first time in a 2D material. Since the phase does not manifest an overall polarization, its detection could be more subtle. As a first step, one may probe such kind of checkerboard-type lattice buckling pattern, e.g., by using scanning tunneling microscopy (STM).

Compared with the 2D FE SnTe-family materials, besides the fundamental distinction (compound vs. elemental), the mechanisms for FE are also different. The SnTe-family materials break the centrosymmetry via the in-plane distortion from the cubic structure as well as the occupation of lattice sites with different kinds of elements. In comparison, the group-V monolayers studied here break centrosymmetry by the out-of-plane atomic-layer buckling, whereas the in-plane distortion does not break the symmetry. As a result, the transformation between two degenerate FE structures of SnTe-family materials requires the breaking and reforming bonds between the atoms [11]; whereas for group-V monolayers, there is no significant bonding and bond-breaking character in the transformation between the two structures A and A'. In addition, the unique structure here permits the presence of an AFE phase (for Bi), which is absent in the SnTe-family materials.

Finally, ferroelectrics should also exhibit piezoelectric behavior. This property can be studied by investigating the polarization change under strain. Interestingly, we find that for Bi, the energy difference between FE and AFE phases can be effectively tuned by strain. In certain range, AFE may even be the ground state with energy lower than FE (see Fig.S12 for a phase diagram). Since strain has been proved to be a powerful tool, especially for 2D materials, this finding provides great opportunity to control the phase and to design a variety of nanoscale devices.

# References:

[1] O. Auciello, J. F. Scott, and R. Ramesh, PHYS TODAY **51**, 22 (1998).
[2] N. Nuraje, and K. Su, NANOSCALE **5**, 8752 (2013).
[3] J. Junquera, and P. Ghosez, NATURE **422**, 506 (2003).
[4] D. D. Fong *et al.*, SCIENCE **304**, 1650 (2004).
[5] I. Kornev, L. Bellaiche, and E. Almahmoud, PHYS REV B **81**, 64105 (2010).
[6] K. Chang *et al.*, SCIENCE **353**, 274 (2016).
[7] B. J. Kooi, and B. Noheda, SCIENCE **353**, 221 (2016).
[8] K. Shportko *et al.*, NAT MATER **7**, 653 (2008).
[9] T. Hu *et al.*, NANO LETT **16**, 8015 (2016).
[10] M. Wu, and X. C. Zeng, NANO LETT **16**, 3236 (2016).
[11] R. Fei, W. Kang, and L. Yang, PHYS REV LETT **117**, 97601 (2016).
[12] M. Bianchi *et al.*, PHYS REV B **85**, 155431 (2012).
[13] T. Nagao *et al.*, PHYS REV LETT **93**, 105501 (2004).
[14] Y. Lu *et al.*, NANO LETT **15**, 80 (2015).
[15] J. Sun *et al.*, PHYS REV LETT **109**, 246804 (2012).

TABLE I: The spontaneous polarization $P_s$ (unit $10^{-10}$ C/m) at zero temperature and fitted parameters in Eq.(1). $A$, $B$, and $C$ describe the double-well potential. $D$ and $E$ are constants representing the mean-field interactions between the first and the second nearest neighbors.

| Element | $P_s$ | $A$ | $B$ | $C$ | $D$ | $E$ |
|---|---|---|---|---|---|---|
| As | 0.36 | -0.0532 | 0.1353 | -0.0782 | 0.049 | 0.075 |
| Sb | 0.75 | -0.1401 | 0.0687 | -0.0084 | 0.043 | 0.050 |
| Bi | 1.67(FE)/1.51(AFE) | -0.0675 | 0.0071 | -0.0002 | 0.008 | 0.019 |

TABLE II: The energy barrier height $E_G$ (meV), Curie temperature $T_c$ (K), and parameters in Eq.(2). The latter three are obtained from fitting the Monte Carlo results.

| | $E_G$ | $T_c$ | $\mu$ | $\delta$ |
|---|---|---|---|---|
| As | 5.83 | 425 | 0.184244 | 0.120454 |
| Sb | 87.5 | 680 | 0.295558 | 0.141036 |
| Bi | 183.6 | 550 | 0.507775 | 0.192268 |

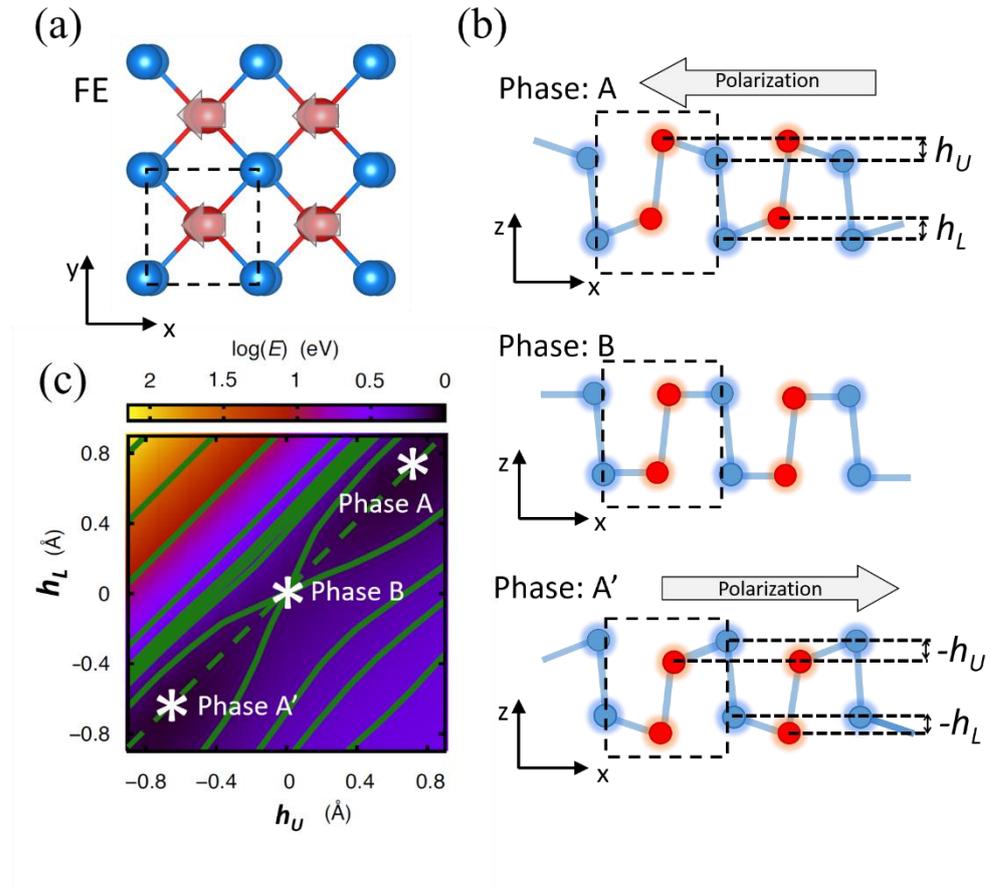

FIG 1 (a) Top view of group-V elemental monolayer. The rectangle with black dashed lines indicates the unit cell. (b) Side views of the two degenerate distorted non-centrosymmetric structures (phases A and A') and undistorted centrosymmetric structure (phase B, corresponding to the phosphorene structure). The height differences between red and blue colored sites in upper and lower atomic layers are labeled as $h_U$ and $h_L$, respectively. (c) Free energy contour for As monolayer versus the buckling heights ($h_U, h_L$). The phases A, A' and B are marked.

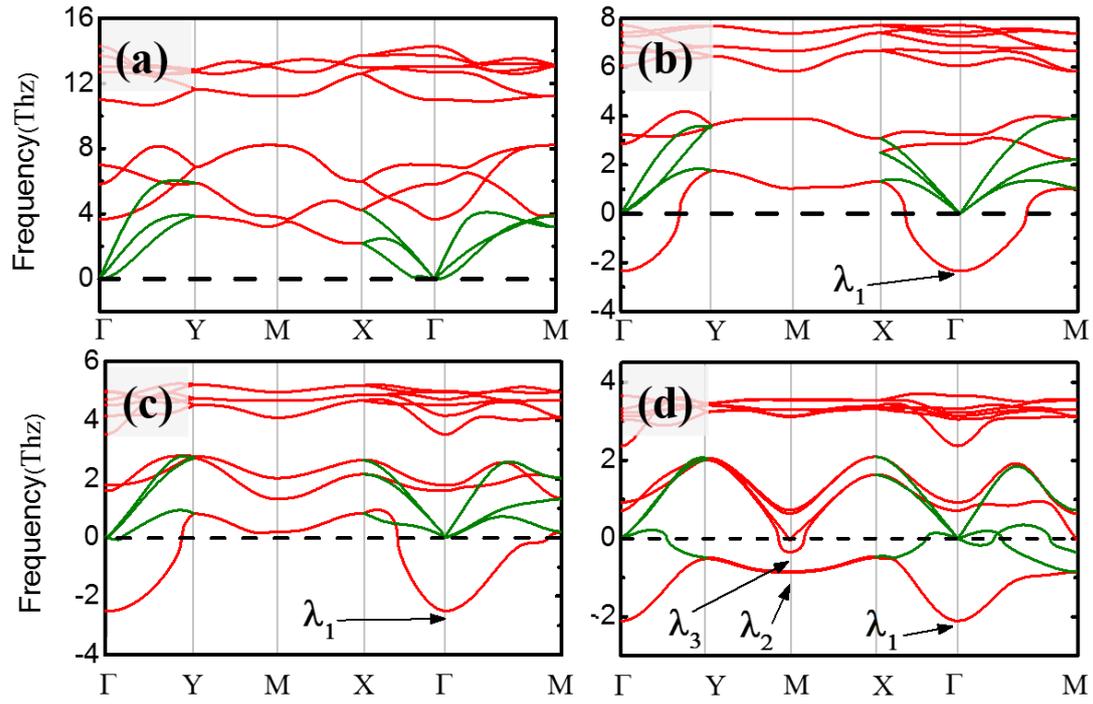

FIG 2 Phonon spectra for group-V monolayers with undistorted centrosymmetric structure (phase B): (a) P, (b) As, (c) Sb and (d) Bi. Green lines mark the acoustic branches while red lines mark the optical branches. $\lambda_1$, $\lambda_2$ and $\lambda_3$ indicate the soft optical phonon modes.

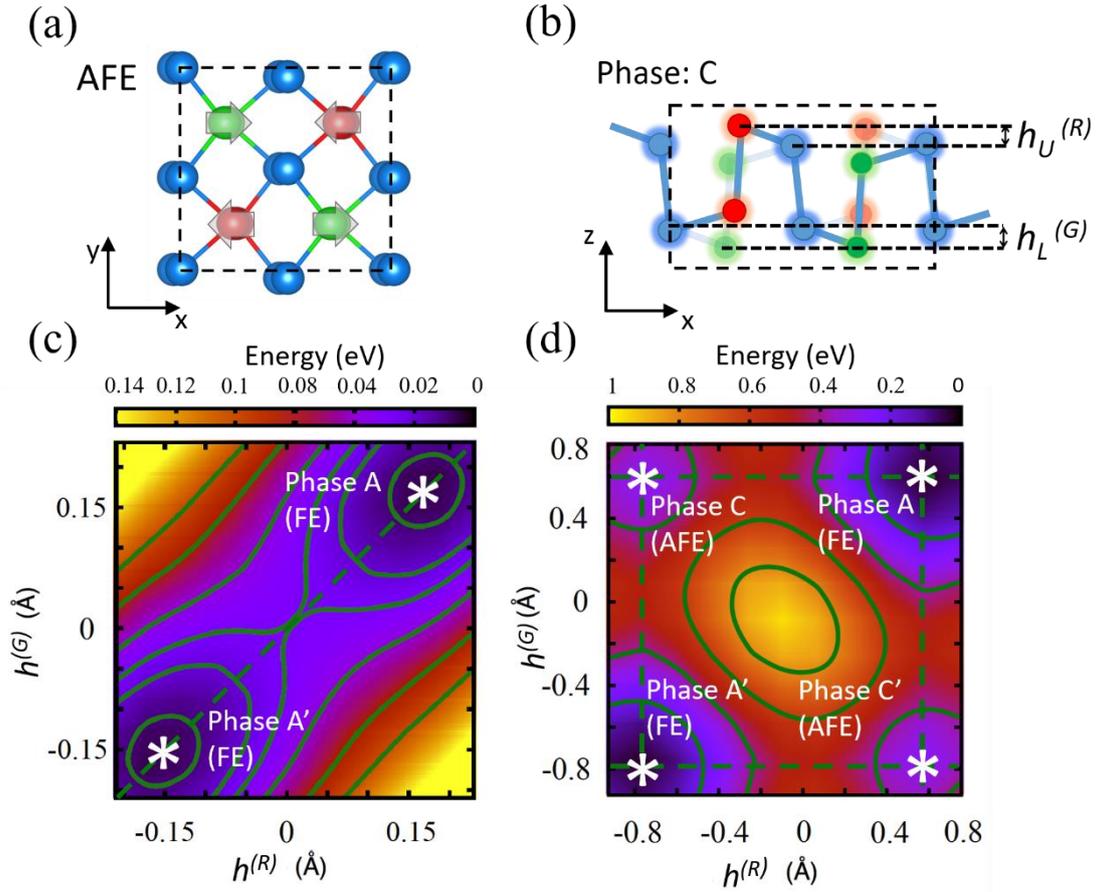

FIG. 3. (a) Top and (b) side view of Bi monolayer in AFE configuration. The atomic sites shifted up and down are colored as red and green, respectively. The blue sites are not shifted in vertical direction. (c, d) Free energy contours of (c) As and (d) Bi monolayers versus $(h^{(R)}, h^{(G)})$. For Bi, besides the FE phases A and A', there also exist metastable AFE phases C and C'.

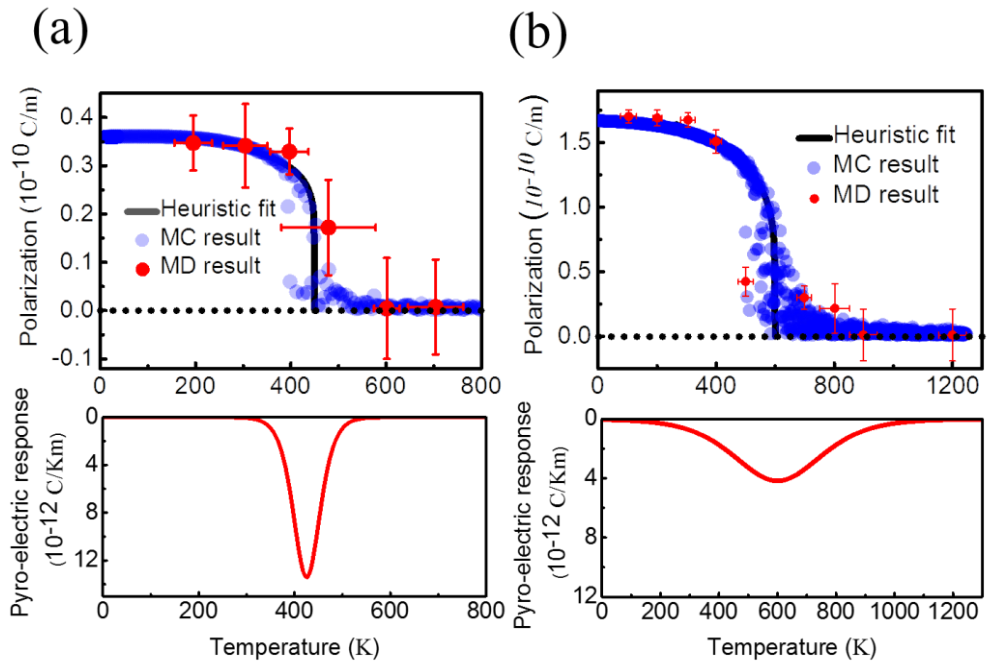

FIG. 4. Upper panel: Temperature dependence of polarization obtained from Monte Carlo simulations (blue dots) and *ab-initio* molecular dynamics (red dots) for (a) As and (b) Bi monolayers. The black lines are heuristic fits of Monte Carlo results. Lower panel: pyroelectric response for As and Bi monolayers.